\documentclass[aps,prl,twocolumn,floatfix,fleqn,10pt,superscriptaddress,showpacs]{revtex4-1}

\usepackage{graphicx}
\usepackage{verbatim}
\usepackage{comment}
\usepackage{amssymb}
\usepackage{enumerate}
\usepackage{psfrag}
\usepackage{pstricks}
\usepackage{color}
\usepackage{ulem}
\emergencystretch=\maxdimen
\newcommand{\Saf}{$S_{\rm{AF}}$~}
\newcommand{\Nl}{N\'eel~}

\newcommand{\lsim}{\mathrel{\hbox{\rlap{\lower.55ex \hbox{$\sim$}} \kern-.3em \raise.4ex \hbox{$<$}}}}

\begin{document}

\title{Magnetic correlations and pairing in the 1/5-depleted square lattice Hubbard model}

\author{Ehsan Khatami}
\affiliation{Department of Physics, University of California, Davis, California 95616, USA}
\affiliation{Department of Physics and Astronomy, San Jose State University, San Jose, California 95192, USA}
\author{Rajiv R. P.~Singh}
\affiliation{Department of Physics, University of California, Davis, California 95616, USA}
\author{Warren E.~Pickett}
\affiliation{Department of Physics, University of California, Davis, California 95616, USA}
\author{Richard T.~Scalettar}
\affiliation{Department of Physics, University of California, Davis, California 95616, USA}

\begin{abstract}
We study the single-orbital Hubbard model on the 1/5-depleted 
square-lattice geometry, which arises in such diverse systems as the
spin-gap magnetic insulator CaV$_4$O$_9$ and ordered-vacancy iron
selenides, presenting new issues regarding the origin of both magnetic
ordering and superconductivity in these materials.
We find a rich phase diagram that includes a plaquette singlet phase, a dimer 
singlet phase, a N\'eel and a block-spin antiferromagnetic phase, and stripe 
phases. Quantum Monte Carlo simulations show that the dominant pairing 
correlations at half filling change character from d-wave in the plaquette 
phase to extended s-wave upon transition to 
the N\'eel phase. These findings have intriguing connections to iron-based 
superconductors, and suggest that some physics of multiorbital systems can 
be captured by a single-orbital model at different dopings.
\end{abstract}


\maketitle

The interplay of magnetic order and pairing correlations has been a
central topic in strongly correlated materials, and, in particular, in
copper-based~\cite{e_dagotto_94} and more recently
iron-based~\cite{r_stewart_11} high-temperature superconductors. That
pairing arises in intimate proximity to magnetism is initially somewhat
surprising, since long-range magnetic order usually leads to an
insulating Mott or Slater gap, which precludes superconductivity.  Much
study of these materials has been devoted to understanding how doping,
and the presence of multiple bands, modify the
magnetism~\cite{FeMag1,FeMag2,FeMag3} and allow pairing and short-range
spin order to complement each other~\cite{c_cruz_08,c_xu_08,a_chubukov_08,q_si_08,g_uhrig_09,FeSup1,r_fernandes_10,FeSup2,k_oakazaki_12,q_ge_13,n_xu_13}.

\begin{figure}[b]
\centerline
{\includegraphics*[width=1.5in]{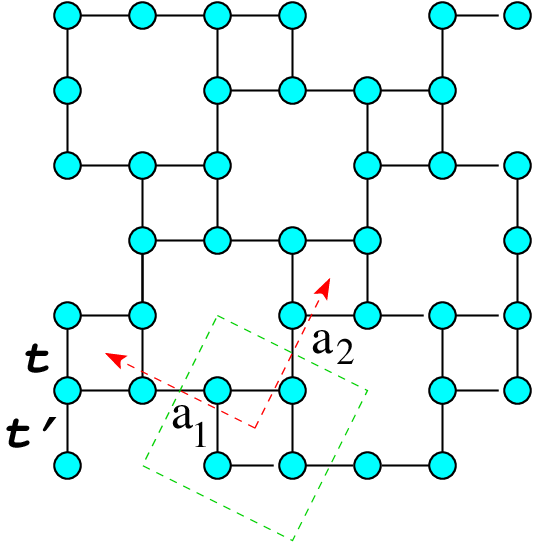}}
\caption{The geometry of the 1/5-depleted square lattice. $2\times2$ plaquettes have 
intersite hopping $t$. Different plaquettes are linked by hopping $t'$. The
two primitive vectors are shown by red arrows and the unit cell is shown by the
tilted square.}
\label{fig:geometry}
\end{figure}

One geometry which has been a recurring structure in real materials,
and which permits tuning of the degree of magnetic order, is the
periodically 1/5-depleted square lattice, consisting of coupled
plaquette unit cells (see Fig.~\ref{fig:geometry})~\cite{n_katoh_95,k_ueda_96,m_troyer_96,m_gelfand_96,cavo_wep}. 
It was first discovered in the study of
spin-gap calcium vanadate material CaV$_4$O$_9$~\cite{s_taniguchi_95}.
More recently, the same structure arises in an ordered vacancy iron
selenide family of materials~\cite{w_bao_11,f_ye_11} where metallic, insulating,
multiple magnetically ordered, and superconducting 
phases arise~\cite{dagotto_rmp,q_ge_13,k_oakazaki_12,n_xu_13,r_thomale_11,s_maiti_11}.

An itinerant Hubbard model in this geometry with a single orbital per site is a four-band model
and can be mapped onto a four plaquette-orbitals model on the nondepleted square lattice.
Such a model allows a systematic exploration of (i) crossovers from weak to strong coupling
behavior, (ii) multiple competing magnetic and spin-gap phases, (iii) possible 
effects of proximity to different phase transitions on the superconducting 
pairing, and (iv) connections between a model with multiple orbitals per site 
and a single-orbital model at different dopings. These make it an important model 
conceptually, and very relevant to the iron selenide family of materials.

Here, we study this single-orbital Hubbard Hamiltonian on the 1/5-depleted
square lattice. Using the determinant quantum Monte Carlo (DQMC) method
\cite{r_blankenbecler_81,supp}, which has no ``minus-sign problem"~\cite{loh90} 
at half filling on this lattice, we find that the dimerized phase of the 
large Hubbard $U$ limit (Heisenberg model) connects smoothly to a band 
insulator as $U$ goes to zero. However, as $U$ decreases, the N\'eel phase 
extends farther and farther into the region where intraplaquette hopping 
is dominant. The plaquette phase, at large $U$, is always separated from 
the metallic phase, obtained at $U=0$, by an intervening N\'eel phase.

In the limit where the interplaquette hopping $t'$ is much smaller than 
the intraplaquette hopping $t$, our model is a variant of the weakly 
coupled plaquette model studied by Tsai and Kivelson~\cite{w_tsai_06}. 
This Hamiltonian can be rigorously shown to have pair binding and a 
superconducting phase at infinitesimal doping away from half filling for 
$U\lesssim4.6t$, a property which remains true for our model as well. 
Our QMC simulations extend the result away from small $t'$ and demonstrate that 
singlet pairing is predominantly in the $d$-wave channel in the plaquette phase 
and becomes particularly large as $U$ exceeds one half of the noninteracting 
bandwidth. We also find that, surprisingly, as soon as one approaches the 
phase transition to the N\'eel order, the dominant pairing changes from 
$d$-wave to extended $s$-wave. To our knowledge, there has been no previous
unbiased demonstration of interchange of superconducting pairing symmetry 
with change in the magnetic properties, emphasizing the 
close interplay of magnetic and superconducting correlations in these 
systems.

Indeed, this observation has possible connections to the iron selenide 
materials whose magnetic phases include ubiquitous stripe phases and a $2\times2$
block-spin antiferromagnet~\cite{w_bao_11,f_ye_11,q_luo_11,h_li_12,dagotto_rmp}.  
In the latter, spins within a plaquette align, and these block spins 
then order in an antiferromagnetic pattern. Using the random
phase approximation (RPA), we have explored a number of different
magnetic instabilities in the 1/5-depleted geometry.  At half filling,
the dominant order in our nearest-neighbor (NN) model is the usual N\'eel
phase.  However, away from half filling, both the stripe phase and 
the $2\times2$ block-spin antiferromagnet are found to be the
leading instabilities over different doping ranges, remarkably, showing that
such phases can arise in models without any frustration or
multiorbital character. 

The Hubbard Hamiltonian considered here is,
\begin{eqnarray}
\hat H=&-&\sum_{ij\sigma}t^{\phantom{\dagger}}_{ij}c^{\dagger}_{i\sigma}
c^{\phantom{\dagger}}_{j\sigma}
-\mu\sum_{i\sigma} n_{i\sigma} \nonumber \\
&+&U\sum_i (n_{i\uparrow}-\frac12)(n_{i\downarrow}-\frac12).
\label{eq:H}
\end{eqnarray}
Here, $c^{\phantom{\dagger}}_{i\sigma}$ 
($c^{\dagger}_{i\sigma}$) annihilates (creates) a
fermion with spin $\sigma$ on site $i$,
$n_{i\sigma}=c^{\dagger}_{i\sigma} c^{\phantom{\dagger}}_{i\sigma}$ is
the number operator, $U$ is the onsite repulsive Coulomb interaction,
and $t_{ij}$ is the hopping matrix element between sites $i$ and $j$.
We allow for NN hopping only and consider two different
values: $t_{ij}=t$ if $i$ and $j$ are
nearest neighbors within a plaquette, and $t_{ij}=t'$ if $i$ and $j$ are
nearest neighbors on a bond that connects two distinct plaquettes. 
At $U=0$, there are four bands  
with dispersion $\epsilon_\alpha(k)$ given
by the roots of, $[\epsilon^2_\alpha(k) - t'^2]^2 - 4 t^2 [ \epsilon_\alpha(k) 
+ t' {\rm cos}\, k_x] [ \epsilon_\alpha(k) + t' {\rm cos}\, k_y]=0$. As we vary 
the ratio $t'/t$, the noninteracting bandwidth $w=4t+2t'$ is kept fixed at 6,
setting the energy unit to $w/6$ throughout the paper.

The richness of the band structure has prompted a recent mean-field study
of the model at quarter filling, where there is on average one half particle 
per site~\cite{y_yamashita_13}. When $t'=t$, the Fermi energy at this 
filling coincides with a Dirac cone structure at the zone center and a flat 
band in its proximity. Yasufumi {\it et al.}~\cite{y_yamashita_13} identify 
three different phases: a paramagnetic insulator, a paramagnetic metal,
and an antiferromagnet, for which phase transitions could be described
by an effective SU(3) theory. The Mott transition in the dimer region
has also been recently studied within a cluster dynamical mean-field 
theory~\cite{y_yanagi_14}.

\begin{figure}[b]
\centerline {\includegraphics*[width=2.5in]{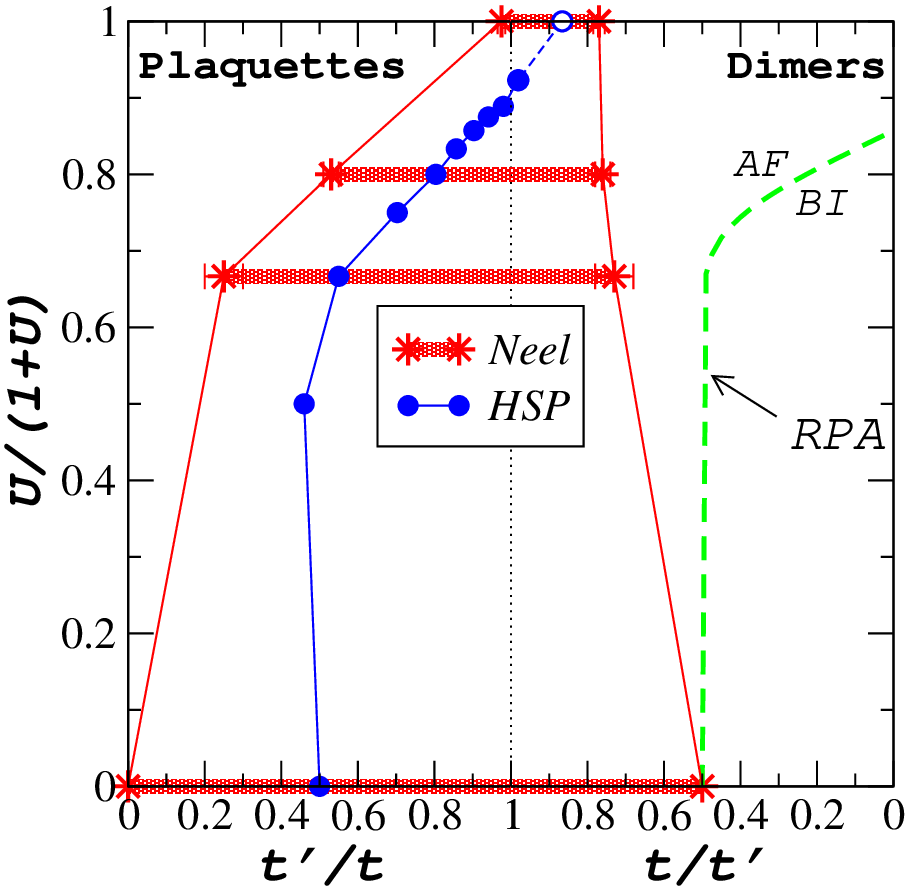}}
\rput(0.2,2.3){\includegraphics[width=0.58in]{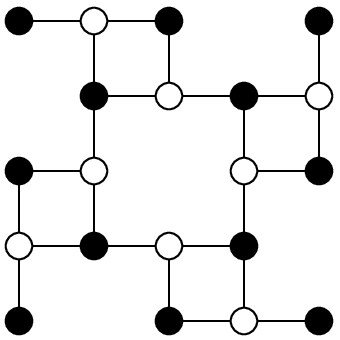}}
\caption{Ground state phase diagram at half filling. The thick horizontal
lines indicate the region with long-range N\'eel order. At $U=\infty$, 
the AF region is obtained from Ref.~\onlinecite{m_troyer_96} (Heisenberg 
model study on the same geometry) by considering $t'/t=\sqrt{J'/J}$ ($J$ 
and $J'$ are the intra- and interplaquette spin exchange interactions). 
Blue circles track the high symmetry point (HSP) inside the AF region 
where the NN spin correlations are equal on all bonds [see
Fig.~\ref{fig:SAF}(a)]. Similar results for the HSP
are not available for the Heisenberg limit. So instead, 
the empty blue circle indicates the location of the maximum of the AF 
moment in that limit. The dashed line shows the AF-band insulator 
(BI) phase boundary as predicted by the RPA. The inset shows the 
AF ordering. Filled (empty) circles denote up (down) spins.}
\label{fig:phase}
\end{figure}

The phase diagram at half filling in the plane of $t'/t$ and $U/(1+U)$ 
is given in Fig.~\ref{fig:phase}. It establishes the dominant magnetic 
instability as antiferromagnetism. The range of $t'/'$ for which the ground
state is N\'eel ordered is shown as thick horizontal lines for three different
values of $U$. At $U\ll 1$, the antiferromagnetic (AF) region extends from an infinitesimal $t'$ 
all the way to $t'/t=2$, beyond which the noninteracting system is a 
band insulator. The N\'eel phase in this regime is favored by AF nesting at
the Fermi surface for $t'<2t$, and the fact that the growing nested
area compensates for the loss of uniformity in the system as $tÕ/t\to 0$.
We obtain this range from the RPA, which is exact in that limit
(the RPA estimate for the AF phase boundary at nonzero 
$U$ is also shown by a dashed line in Fig.~\ref{fig:phase}). 

As we turn on the interaction, we find that for $t$ and $t'$ 
sufficiently close to each other, there is always a
nonzero N\'eel order parameter in the thermodynamic limit. 
We locate the phase boundary by finite-size scaling of the DQMC AF 
structure factor, $S_{\rm AF}$~\cite{supp}. One can see that as $U$ increases,
the N\'eel ordered region shrinks, especially on the plaquette side,
and moves to the Heisenberg limit ($U\to\infty$) range~\cite{m_troyer_96}.

Also shown as filled circles in the phase diagram of Fig.~\ref{fig:phase}
are the hopping ratios at which the intra- and interplaquette NN spin
correlations are equal in magnitude. This line of ``high symmetry points"
(HSPs) favors the plaquette side of the phase diagram until it
veers toward the dimer side around $U$=3, tracking the magnetically 
ordered region.

\begin{figure}[b]
\centerline {\includegraphics*[width=3.3in]{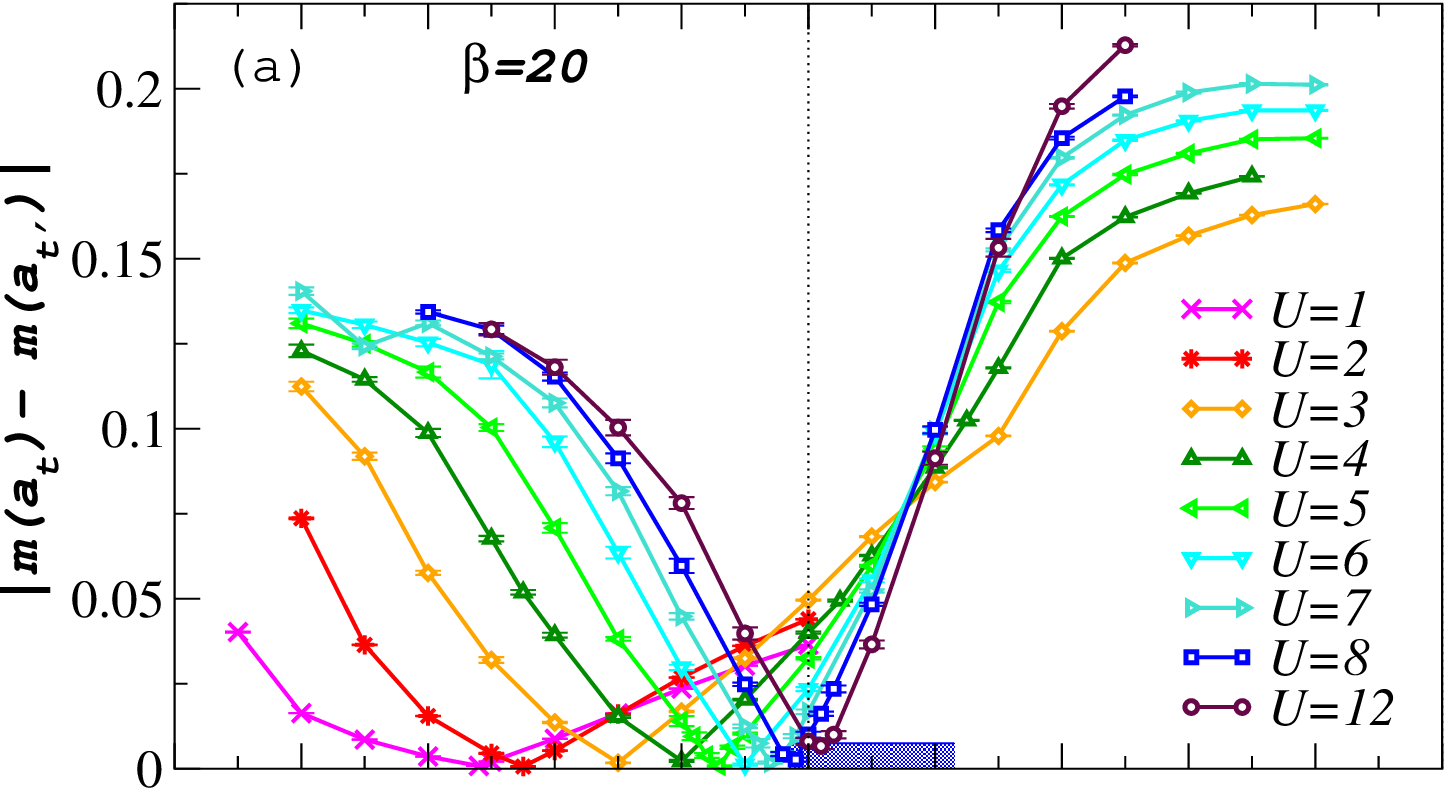}}
\centerline {\includegraphics*[width=3.3in]{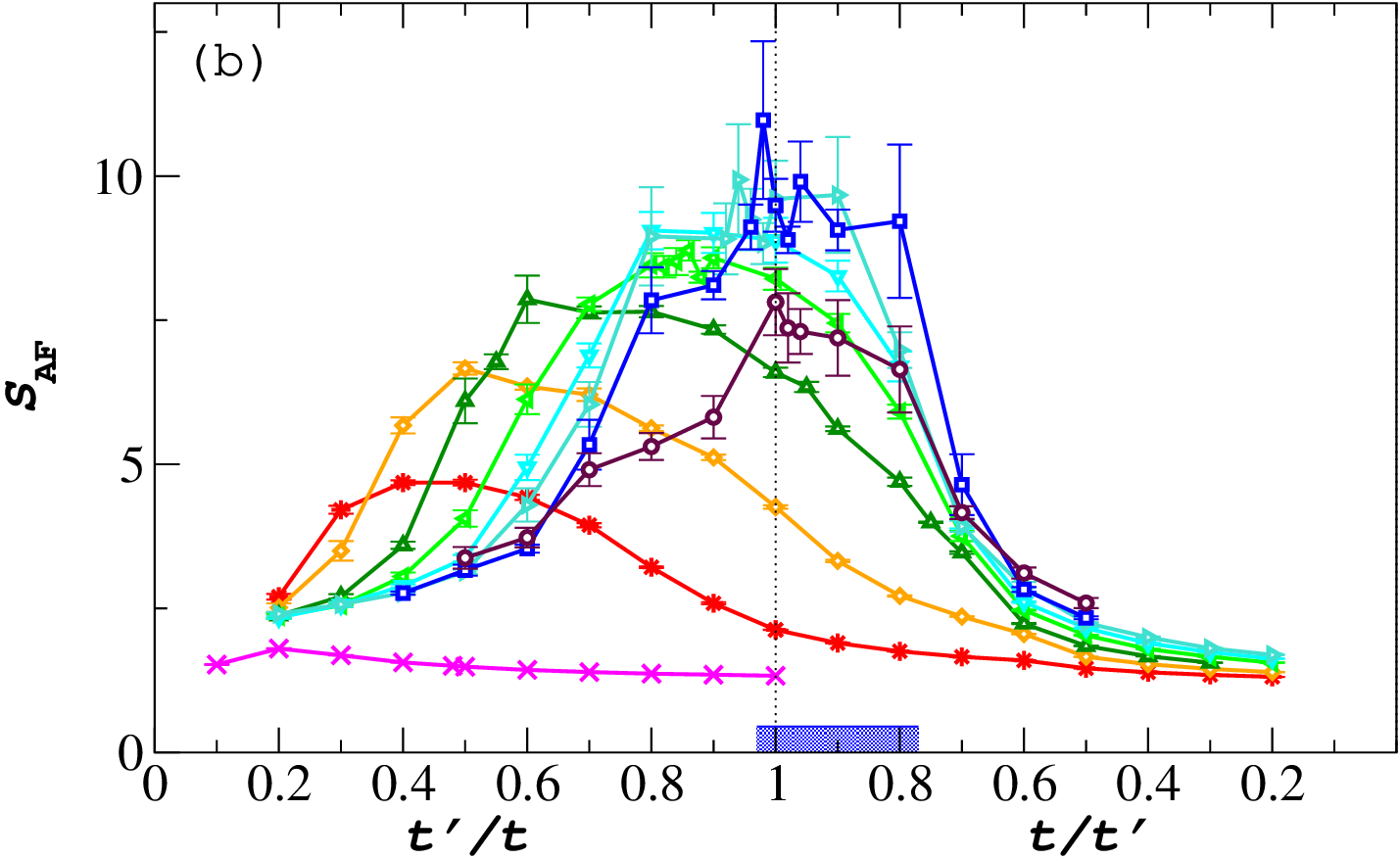}}
\caption{(a) The absolute value of the difference in the NN spin
correlations on $t$ and $t'$ bonds at $\beta=20$ from DQMC 
vs $t'/t$ for several values of the interaction strength [$m({\bf r})$ is
the spin-spin correlation function at distance ${\bf r}$ and ``a" denotes 
the lattice constant between NNs~\cite{supp}]. The shaded region 
on the horizontal axis is added to show the boundaries of the N\'eel phase 
in the $U\to \infty$ limit. The lattice is a $4\times4$ arrangement of $2\times 2$ plaquettes
($N=64$), except for $U=1$ and 2 where the $8\times8$ arrangement ($N=256$)
is used. We have also simulated a 576-site lattice for the latter interactions 
and found no significant changes in the location of the HSP.
(b) The AF structure factor vs $t'/t$ at $\beta=20$ from DQMC. Except for 
$U=1$, for which $N=256$, the results are obtained for the $N=64$ lattice. }
\label{fig:SAF}
\end{figure}

Figure \ref{fig:SAF}(a) shows the absolute value of the difference of 
NN spin correlations on the two types of bonds at inverse temperature 
$\beta=20$ as a function of hopping ratios. At the weakest 
coupling $U=1$ the NN spin correlation on the intraplaquette $t$ 
bonds exceeds the interplaquette $t'$ bonds up to $t'/t \sim 0.5$, 
at which point the relative size is reversed~\cite{note2}. However, at 
the strongest coupling studied, $U=12$, the intraplaquette spin correlation 
remains larger all the way to $t'/t \sim 1$. The finite-size 
dependence of these correlations is either negligible, or has 
been taken into account~\cite{supp} (see caption of Fig.~\ref{fig:SAF} 
for details). We note that all of the calculated NN spin correlations 
are antiferromagnetic, regardless of the value of $t'/t$ or $U$.

The results in Fig.~\ref{fig:SAF}(b) show the low-temperature 
$S_{\rm AF}$ as a function of the hopping ratios for the 
same range of interaction strengths as in Fig.~\ref{fig:SAF}(a).
Although these results are for a single (relatively large) lattice size, 
the evolution of the peak of $S_{\rm AF}$ clearly conveys the trend 
in the long-range order as $U$ is increased towards the Heisenberg limit.
These maxima shift steadily from the plaquette side at weak coupling 
to the dimer side at strong coupling.

\begin{figure}[b]
\centerline
{\includegraphics*[width=3.2in]{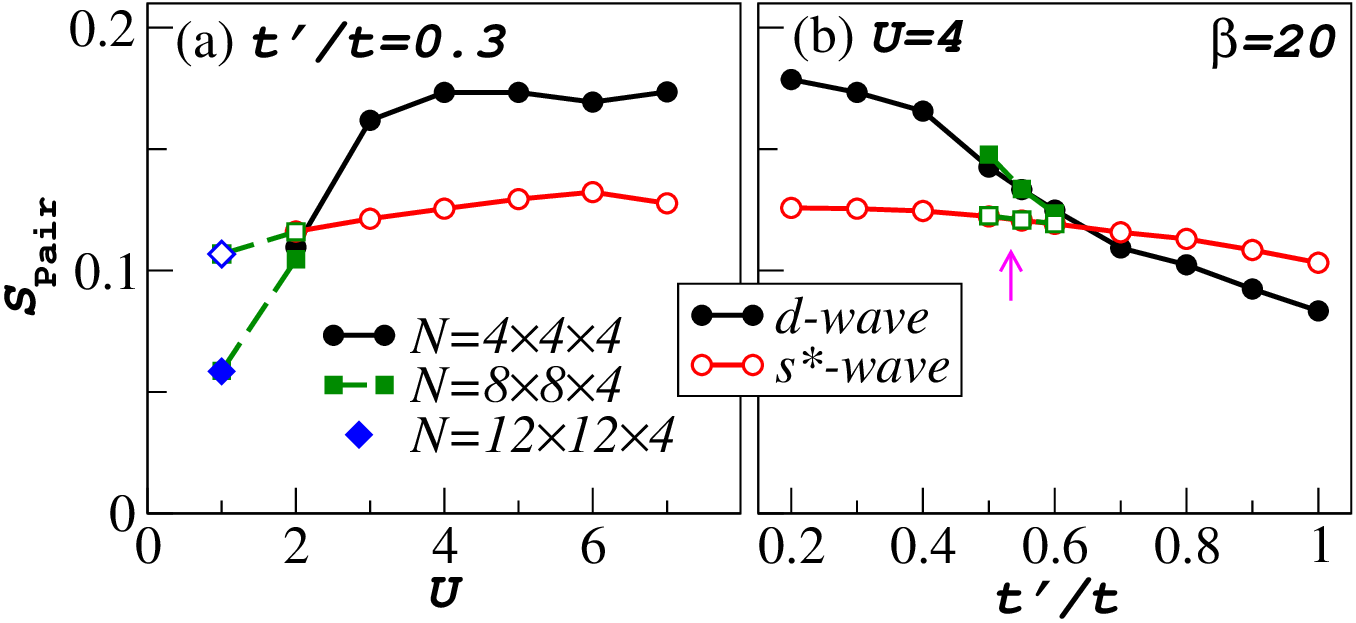}}
\caption{(a) Pairing structure factor~\cite{supp} at $\beta=20$ and $t'/t=0.3$ vs the 
interaction strength. For $U=1$ and 2, two different system sizes are shown. 
(b) Pairing structure factor at $\beta=20$ and $U=4$ vs the ratio $t'/t$. 
Full (empty) symbols are for the d-wave (extended s-wave) symmetry. The 
error bars are smaller than the symbols. The arrow indicates the location
of the AF phase transition.}
\label{fig:SCStruc}
\end{figure}

An intriguing feature seen in this model is the change in symmetry of  
low-temperature pairing correlations from d-wave in the plaquette phase 
to extended s-wave upon entering the N\'eel phase. This is demonstrated 
in Fig.~\ref{fig:SCStruc}, where we plot the uniform pairing structure
factor~\cite{supp} for the two symmetries versus the interaction strength at $t'/t=0.3$,
and vs the hopping ratio at $U=4$, for which we know the location of the AF phase
transitions. As shown in Fig.~\ref{fig:SCStruc}, finite-size effects at small $U$
are not responsible for this difference. We have also verified that the
values of the structure factor do not change significantly by further lowering the 
temperature. At $U=4$, the change in the pairing symmetry takes place inside
the AF region just before the transition to the plaquette phase. For all
the other interaction strengths, the location of this crossover appears to 
fall to the right (larger $t'$ side) of the AF phase boundary. 
As the charge gap is nonzero in both the AF and the 
plaquette phase, we do not expect to find superconductivity at half 
filling. However, the strength of the pairing at half filling should be
indicative of the nature of superconductivity upon doping. The d-wave pairing
in the weakly coupled plaquette phase agrees with the general arguments of Scalapino
and Trugman~\cite{scalapino_trugman} and of Tsai and Kivelson~\cite{w_tsai_06}. 
The dominance of extended s-wave pairing near the
phase transition is a surprising result and points to the close interplay between
magnetism and superconductivity in these systems.

\begin{figure}[t]
\centerline
{\includegraphics*[width=3.2in]{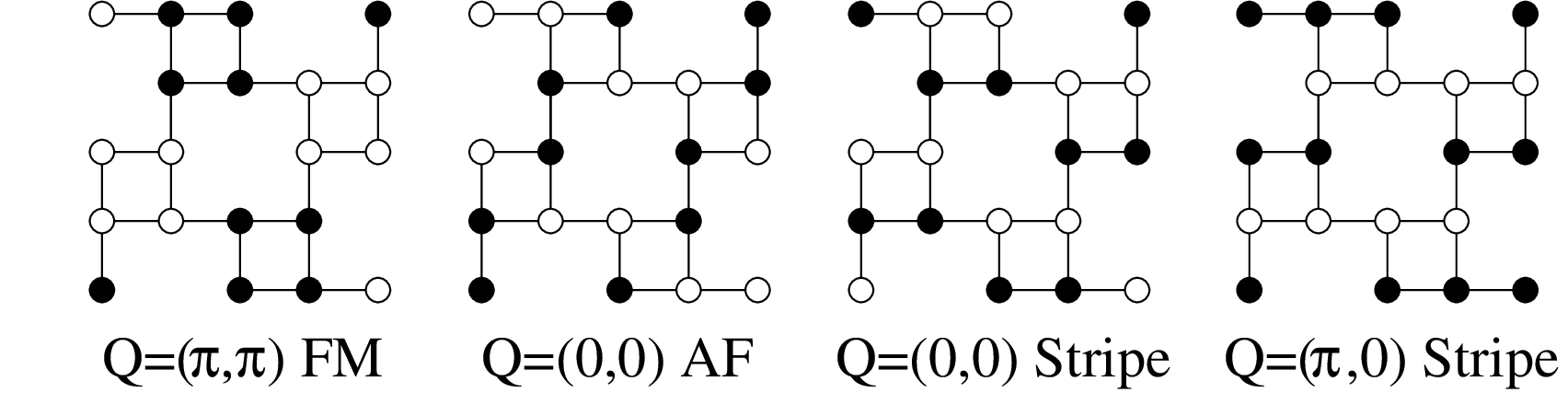}}
\centerline {\includegraphics*[width=3.3in]{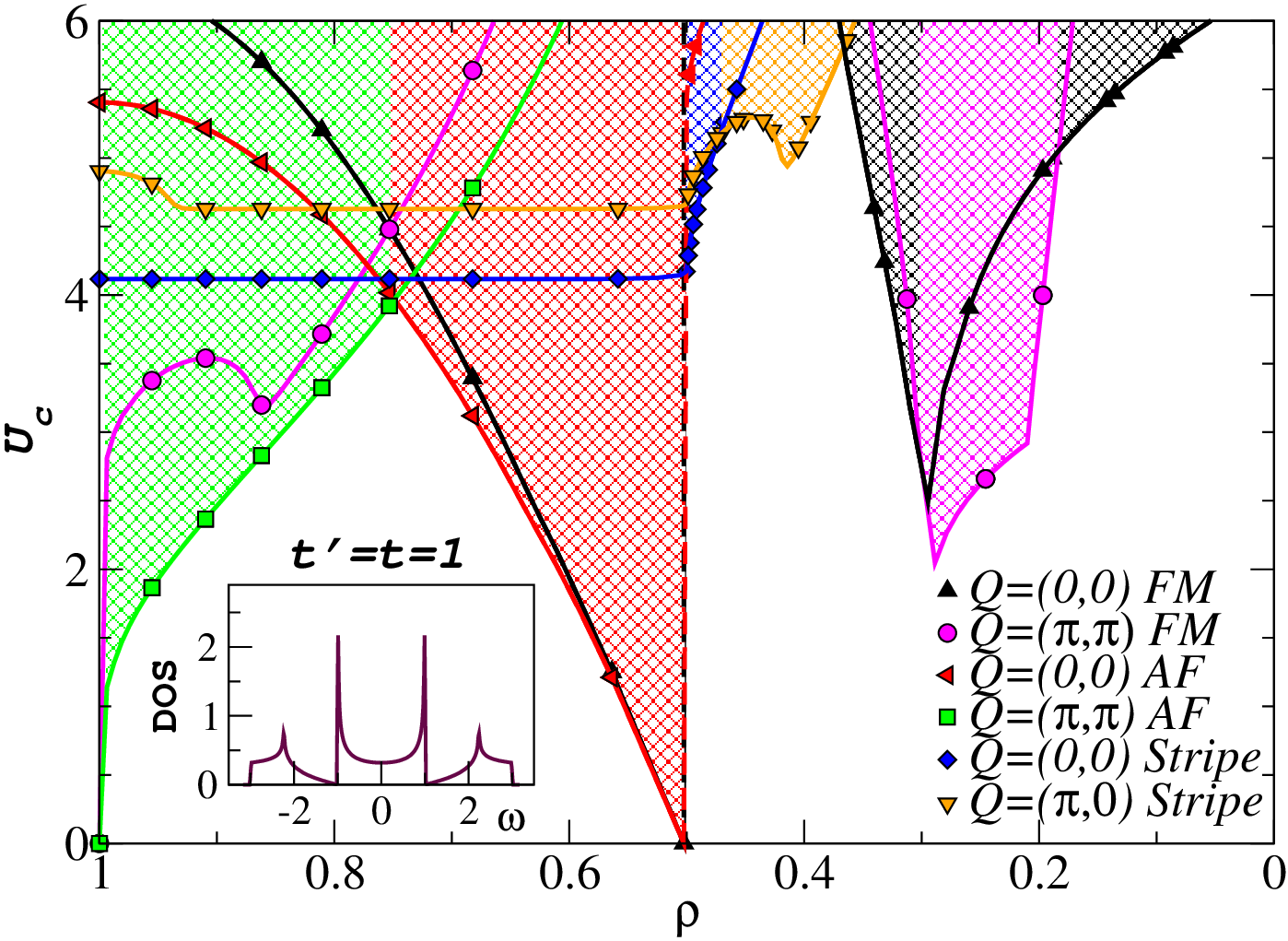}}
\caption{(Top) Four of the magnetic orderings considered in this study. The up 
(down) spins are denoted by filled (empty) circles. $Q$ indicates the phase 
between unit cells and the FM, AF, and stripe denote the ordering of spins within 
each unit cell. (Bottom) The ground state RPA phase diagram of the model at 
$t'=t$ away from half filling. The inset shows the density of states (DOS).}
\label{fig:rpa2}
\end{figure}

We now turn to the case away from half filling, where we use the RPA 
instead of the DQMC method, as low-temperature results are not available 
for the latter. The RPA, which is reasonably accurate only at weak couplings, can 
offer insight into the competition between different magnetic orderings
that this  geometry may favor in different doping regions. Figure \ref{fig:rpa2} 
provides the full evolution of the critical interaction strength $U_c$
for six different magnetic orderings as a function of the electron density
$\rho$ in the uniform $t=t'$ case. Four of the magnetic phases are 
shown atop the main panel in Fig.~\ref{fig:rpa2}. The other two are 
the regular $Q=(\pi,\pi)$ AF (shown in Fig.~1)
and the simple $Q=(0,0)$ ferromagnetic (FM) phases. Here, $Q$ is
the wave vector corresponding to the superlattice and the following
letters describe the order within a plaquette. The $Q=(\pi,\pi)$ AF
dominates near half filling and up to $\rho \sim 0.75$.  At that point,
the $Q=(0,0)$ AF has the largest susceptibility and hence, the smallest
$U_c$, even though for densities close to, but higher than $\rho=0.5$ 
(quarter filling), it is degenerate with the $Q=(0,0)$ FM phase. 
Exactly at quarter filling, we find that the $Q=(0,0)$ stripe, and 
not the $Q=(0,0)$ AF, as predicted by a previous mean-field 
calculation~\cite{y_yamashita_13}, is the dominant order. However, 
at a slightly smaller $\rho$, this order is replaced by the $Q=(\pi,0)$ 
stripe order. Interestingly, at $\rho\sim0.2$ to 0.3, the block AF phase, 
observed in ordered-vacancy iron selenide materials, has the lowest 
$U_c$. This order shows up at even lower energies in the anisotropic 
case of $t'<t$. Thus, in many ways, the single-orbital model at
different dopings captures the richness of the magnetic phases observed 
in the iron pnictide and chalcogenide family of materials.

Quantum Monte Carlo methods allow for an exact treatment of 
the combined effects of correlation and band structure on 
lattices of finite spatial size, or equivalently, with finite
resolution in momentum space. Previous DQMC studies of
the effect of multiple bands and different intersite hoppings
on magnetic order have mostly been confined to 
layered geometries in which two spatially extended regions each with
a unique hopping are coupled~\cite{r_scalettar_94,a_euverte_12,a_euverte_13}.
Here, in contrast, we have presented results for a hopping
pattern in which two different $t_{ij}$ are mixed locally, and found 
that tuning their ratio leads to multiple quantum phase transitions
and rich phase diagrams. 
We have also studied the superconducting properties of our model 
at half filling within the DQMC method. Remarkably, the dominant pairing 
symmetry changes character from d-wave in the plaquette phase to
an extended s-wave in the N\'eel phase, revealing an interesting 
interplay between magnetic and superconducting correlations. 
Although our system is insulating at half filling, the dominant 
pairing at half filling should be an indicator of the nature of 
superconductivity upon doping. Moreover, the behavior of both the magnetic 
and superconducting correlations in our single-orbital model offer surprising 
connections to iron-based superconductors, which are multiorbital systems,
implying that they can be mapped to effective one-orbital
models but with varying doping values.

This work was supported by the Department of Energy under Grant No. DE-NA0001842-0 
(E. K. and R. T. S.) and by the National Science Foundation Grants No. PIF-1005503 (R. T. S.), 
No. DMR-1004231 and No. DMR-1306048 
(E. K. and R. R. P. S.) and No. DMR-1207622 (W. E. P.). This work used the Extreme Science 
and Engineering Discovery Environment under Project No. 
TG-DMR130143, which is supported by NSF Grant No. ACI-1053575.
We thank D. Chicks for additional useful input.

\newpage
\
\newpage

\onecolumngrid


\begin{center}

{\large \bf Supplementary Materials:
\\ Magnetic correlations and pairing in the 1/5-depleted square lattice Hubbard model}\\

\vspace{0.6cm}

Ehsan Khatami$^{1,2}$, Rajiv R.P.~Singh$^1$, Warren E.~Pickett$^1$, Richard T.~Scalettar$^1$\\
{\it Department of Physics, University of California, Davis, California 95616, USA}\\
{\it Department of Physics and Astronomy, San Jose State University, San Jose, California 95192, USA}\\

\end{center}

\vspace{1.cm}

\twocolumngrid

\section{Determinantal quantum Monte Carlo}

In the determinantal quantum Monte Carlo (DQMC)~\cite{r_blankenbecler_81,note}, the exponential 
${\rm exp}(-\beta \hat H)$ in the partition function 
is written as a product of incremental imaginary time propagators ${\rm exp}
(-\Delta \tau \hat H)$ where $\beta = L \Delta \tau$ is the inverse temperature.  
The Trotter approximation ${\rm exp}(-\Delta \tau \hat H) \approx {\rm
exp}(-\Delta \tau \hat K) {\rm exp}(-\Delta \tau \hat V)$ is used to
isolate the exponential of the on-site interaction $\hat V$ from the
kinetic energy and chemical potential terms $\hat K$.

A (discrete) Hubbard-Stratonovich (HS) variable $s_{i\tau}$ is
introduced at each spatial site $i$ and imaginary time slice $\tau$ to
decouple the interaction term,
\begin{eqnarray}
e^{-U\Delta \tau (n_{i \uparrow} - \frac12)
(n_{i \downarrow} - \frac12) } &=&  \nonumber \\
\frac12 e^{-U \Delta \tau/4} \sum_{s_{i\tau}=\pm 1}
&e^{ \lambda s_{i \tau} (n_{i \uparrow} - n_{i \downarrow})}&
\end{eqnarray}
where ${\rm cosh}\, \lambda = e^{U \Delta \tau/2}$.  The quartic term in
fermion creation and destruction operators is thereby replaced by quadratic 
terms coupled to the HS field $\{s_{i\tau} \}$.  The fermion degrees of 
freedom can be traced out analytically, leaving a sum over  $\{s_{i\tau} \}$ 
to be performed stochastically (using Monte Carlo).

The result of tracing over the fermion degrees of freedom is a product of two
determinants, one for each spin species.  In general, this product can go
negative, precluding its use as a Monte Carlo weight.  This is known as
`the sign problem' \cite{loh90}, and results in a limit on the
temperatures accessible to the simulation.  In special cases, such as at
half filling ($\mu=0$) and with a bipartite lattice, the signs of the
two determinants are always equal and it is possible to make $\beta$
very large.  However, at generic fillings, the constraint is typically $\beta
t \lesssim 4$ (although the precise value depends on the average density $\rho$, 
the interaction strength $U$, and the lattice size).  For this reason, 
we focus on half filling (which is, anyway, also the density for which the 
mapping to the Heisenberg model is valid), and use the analytic RPA treatment 
to discuss doped lattices. In the work reported here, we use $\Delta \tau = 
1/2U$, except for $U=1$ where $\Delta \tau =1/4$, to keep the associated 
Trotter errors smaller than statistical ones from the Monte Carlo sampling 
of the HS field.

Magnetic properties of the model are determined by the
real space spin-spin correlation functions
\begin{eqnarray}
m^{xx}({\bf r}) &=& \langle \, S^-(i+{\bf r}) S^+(i) \, \rangle \nonumber \\
m^{zz}({\bf r}) &=& \langle \, S^z(i+{\bf r}) S^z(i) \, \rangle,
\end{eqnarray}
where
\begin{eqnarray}
S^+(i) &=& c^{\phantom{\dagger}}_{i\downarrow} c^{\dagger}_{i\uparrow} \nonumber \\ 
S^-(i) &=& c^{\phantom{\dagger}}_{i\uparrow} c^{\dagger}_{i\downarrow}\nonumber \\
S^z(i) &=& \frac12 (n_{i \uparrow} - n_{i \downarrow}),
\label{eq:realspace}
\end{eqnarray}
their average along an arbitrary direction
\begin{eqnarray}
m({\bf r}) = \frac13 [2 m^{xx}({\bf r})+m^{zz}({\bf r})],
\end{eqnarray}
and its Fourier transform, the magnetic structure factor,
\begin{eqnarray}
S({\bf q}) = \sum_{{\bf r}} e^{i {\bf q\cdot r}} m({\bf r}).
\label{eq:structurefactor}
\end{eqnarray}
When long-range order with wavevector ${\bf q}$ is present in $m({\bf r})$, the
spatial sum in Eq.~\ref{eq:structurefactor} at the corresponding ${\bf q}$
diverges in the thermodynamic limit, since the sum
over ${\bf r}$ grows linearly with $N$.  Spin wave theory~\cite{d_huse_88}
predicts that the finite-size correction is linear in the inverse
lattice size $L$, so that $S({\bf q})/N = A + B/L$, where $A$ is proportional to
the square of the magnetic order parameter.  In the disordered phase,
the sum cuts off and $S({\bf q})/N$ is proportional to $1/N = 1/L^2$. Note that 
here, ${\bf q}$ denotes both the phase between sites of a unit cell,
and ${\bf Q}$, the wavevector describing the phase between unit cells.

We consider six possibilities for the spatial structure of the magnetic 
order. Four of them are 
illustrated in Fig. 4 of the main text. The other two are the regular 
antiferromagnetic (AF) order, illustrated in Fig. 1 of the main text, where the 
sign of the spin alternates between nearest-neighbor (NN) sites and the FM in 
which all spins are pointing to the same direction.

\begin{figure}[t]
\centerline {\includegraphics*[width=3.3in]{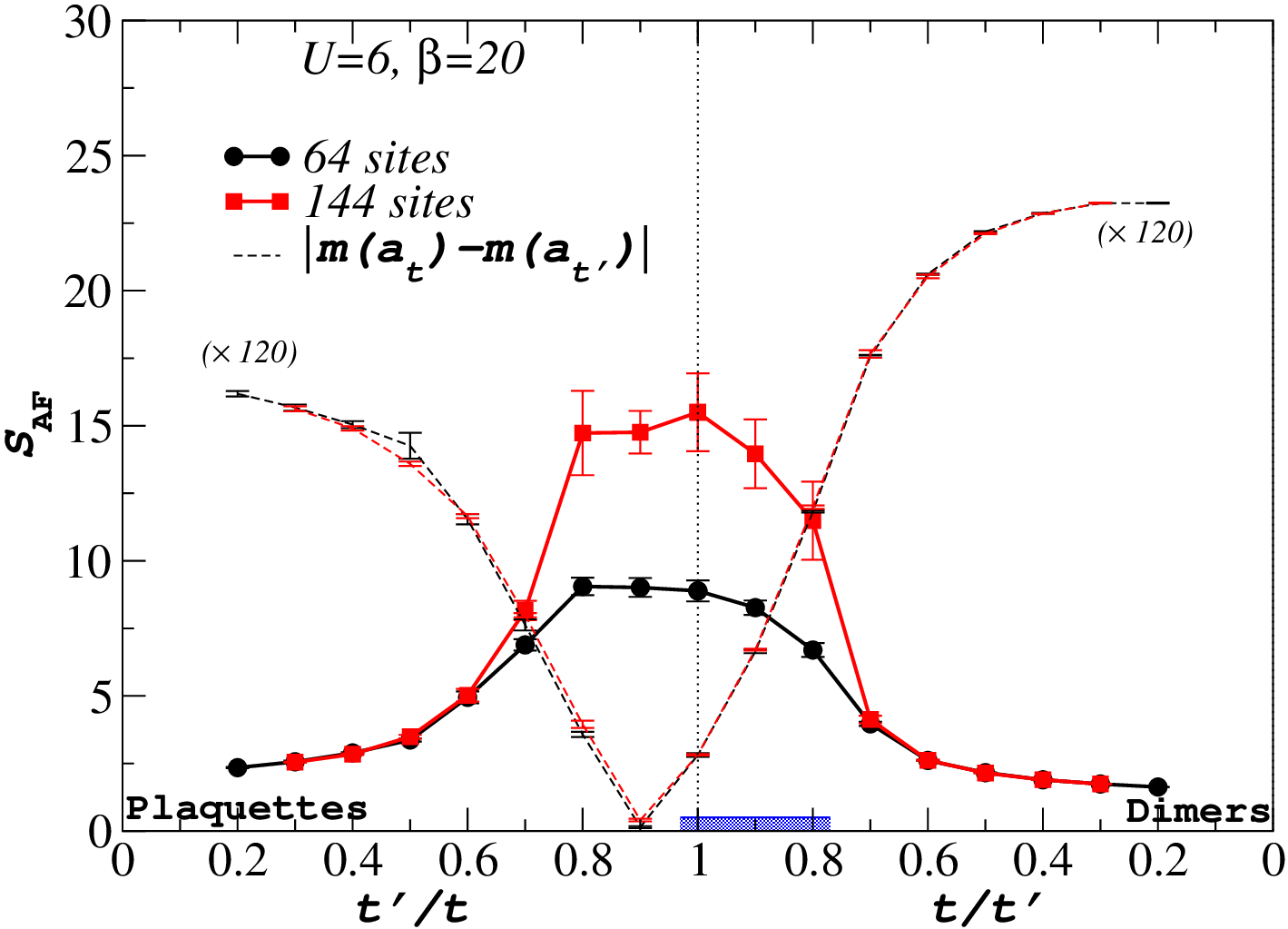}}
\caption{The AF structure factor \Saf for $U=6$ at $\beta=20$ vs $t'/t$ for 
two different system sizes. The difference in the NN spin correlations for 
the same parameters is also shown for the two sizes. The blue (shaded) region
indicates the range of $t'/t$ for which the infinite-$U$ limit exhibits 
long-range AF order.}
\label{fig:U6}
\end{figure}

The magnetic susceptibility can also be evaluated in DQMC by considering
spin correlation functions in real time and separated by
imaginary time $\tau$, e.g., 
\begin{eqnarray}
m^{xx}({\bf r},\tau) &=& \langle \, S^-(i+{\bf r},\tau) S^+(i,0) \, \rangle
\nonumber \\
S^+(i,\tau) &=& e^{\tau \hat H } c^{\phantom{\dagger}}_{i\downarrow}
c^{\dagger}_{i\uparrow} e^{-\tau \hat H },
\end{eqnarray}
and integrating over $\tau$,
\begin{eqnarray}
\chi({\bf q}) &=& \int_0^\beta \, d \tau \sum_{{\bf r}} e^{i {\bf q\cdot r}} m({\bf r},\tau).
\label{eq:realspaceandtime}
\end{eqnarray}
However, for the determination of the phase diagram, the equal time structure 
factor is sufficient.

Similarly, one can define the equal-time pairing structure factor as 
\begin{eqnarray}
S_{\rm Pair}({\bf q}) = \sum_{{\bf r}} e^{i {\bf q\cdot r}} C({\bf r}),
\label{eq:pair}
\end{eqnarray}
where 
\begin{eqnarray}
C({\bf r}) &=& \langle \, \Delta_{\alpha}^{\dagger}(i+{\bf r}) \Delta_{\alpha}(i)
+\Delta_{\alpha}(i+{\bf r}) \Delta_{\alpha}^{\dagger}(i)\rangle
\end{eqnarray}
is the pair-pair correlation function. Here, the pairing operator for the 
symmetry $\alpha$ is defined as
\begin{equation}
\Delta_{\alpha}(i)= \sum_{j\ {\rm NN\ of}\ i}f_{\alpha}(j)
(c_{i\uparrow}c_{j\downarrow}-c_{i\downarrow}c_{j\uparrow}).
\end{equation}
In case of the extended s-wave symmetry, $f_{\alpha}(j)$ is $+1$ for all 
of the three NN of $i$, and for the d-wave symmetry, $f_{\alpha}(j)$ is $+1$
if $j$ is a NN of $i$ along the x axis and $-1$ if along the y axis. We consider 
$S_{\rm Pair}({\bf q}=0)$ as the pairing structure factor, which is shown
in Fig. 3 of the main text.

\section{Antiferromagnetic Phase Transitions}

Figure~\ref{fig:U6} gives additional insight into the finite size 
scaling of the AF structure factors shown in Fig. 2(b) of the main 
text by exhibiting its behavior at $U=6$ with changing the hopping 
ratio for two lattice sizes. For a range of values within the peak,
$S_{\rm AF}$ is not only large, but also grows significantly
with lattice size, giving rise to the nonzero $1/L \rightarrow 0$
extrapolations discussed below. We are also plotting in Fig.~\ref{fig:U6}
the NN spin correlation difference for the two different system sizes, 
which supports our claim that the finite-size corrections for this 
quantity are generally negligible.

\begin{figure}[t]
\centerline {\includegraphics*[width=3.3in]{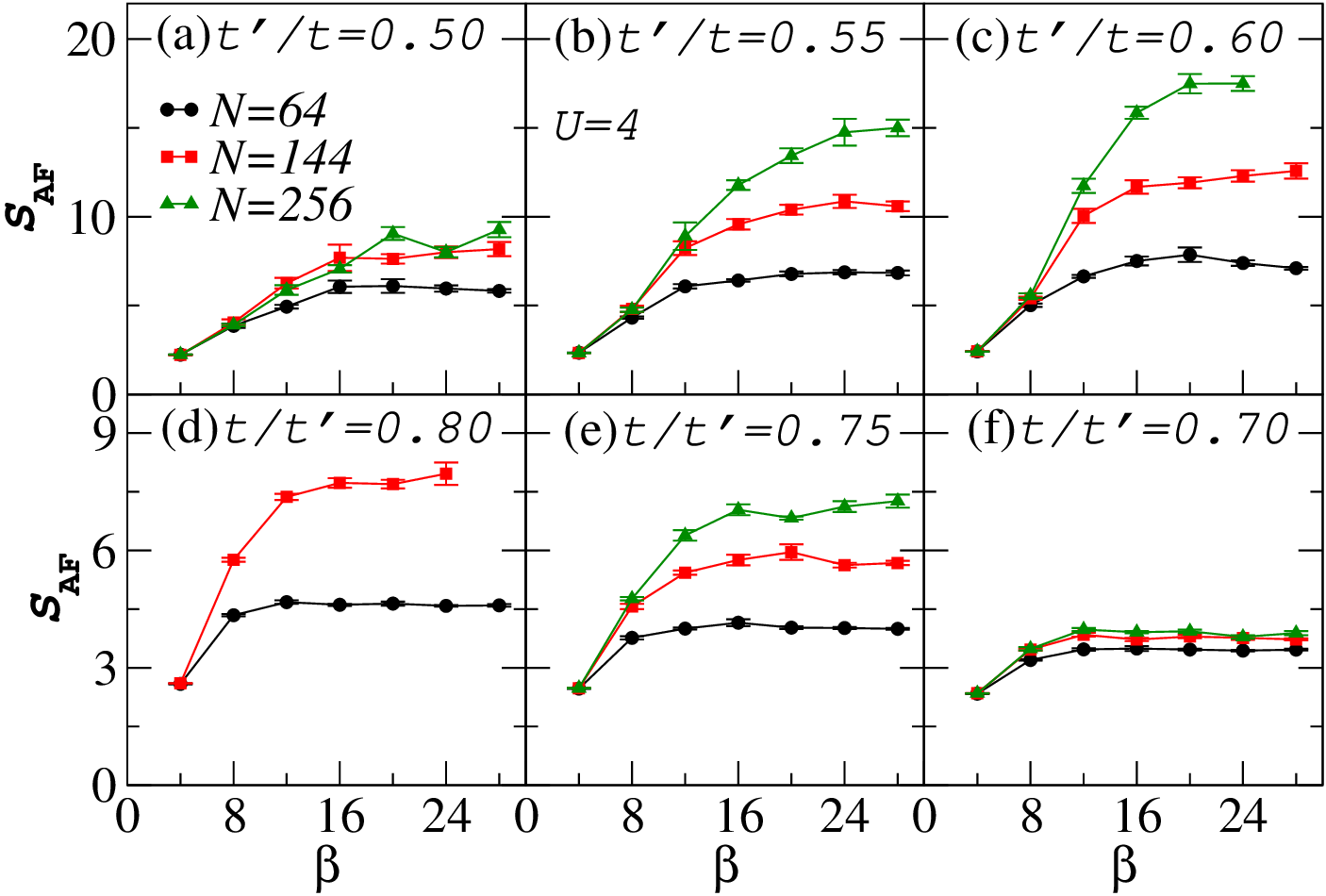}}
\caption{The AF structure factor \Saf vs inverse temperature for several 
values of $t'/t$ and three system sizes. For most cases, \Saf plateaus 
at or below $\beta=20$.}
\label{fig:Safvsbeta}
\end{figure}

\begin{figure}[t]
\centerline {\includegraphics*[width=3.in]{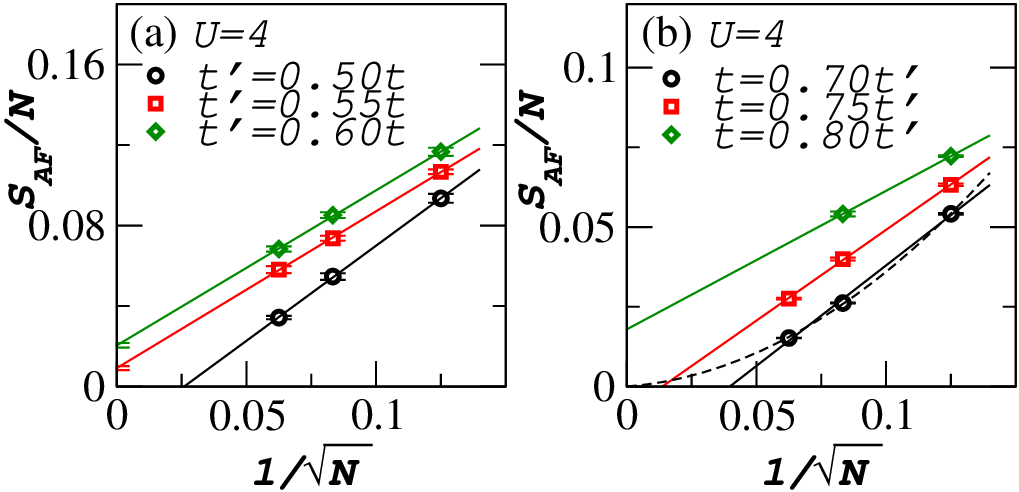}}
\caption{Normalized AF structure factor vs the linear size of 
the system at $U=4$ and for several ratios of $t'/t$. The extrapolated values 
in the thermodynamic limit are used to estimate the \Nl phase boundary at this 
interaction strength. Solid lines are linear fits and the dashed line in (b) 
is a parabolic fit that passes through the origin.}
\label{fig:scaling}
\end{figure}

In order to make sure that $\beta=20$ is a low enough temperature
to describe the ground state properties of the lattice sizes 
we have considered, we show in Fig.~\ref{fig:Safvsbeta} the dependence 
of \Saf at $U=4$ on the inverse temperature across the two transition 
points to the \Nl phase. For each of the ratios of $t'/t$, 
\Saf saturates to a lattice size dependent value. The mean values 
after saturation are used to perform the extrapolations in Figs.
~\ref{fig:scaling}(a) and \ref{fig:scaling}(b). In almost all cases, 
the AF structure factor plateaus at or below $\beta=20$. We have
used the $4\times4$, $6\times6$, and $8\times8$ arrangments of the 4-site
unit cell as our clusters for the extrapolations, corresponding to $N=64$, 144, 
and 256, respectively.

As discussed above, in the N\'eel phase, the normalized structure factor approaches a 
finite value in the thermodynamic limit as a function of the inverse 
linear size of the system. In the disordered phases, one expects 
$S_{\rm AF}/N$ to vanish as $1/N$. Consequently, after proper fitting 
of the data, the two end points of the N\'eel region at $U=4$ are
estimated from the extrapolations of $S_{\rm AF}$ to the thermodynamic 
limit for several values of the $t'/t$ in their proximity as shown in
Figs.~\ref{fig:scaling}(a) and \ref{fig:scaling}(b).

\section{Random Phase Approximation}

In the RPA method, the non-interacting magnetic susceptibility is
evaluated via,
\begin{eqnarray}
\chi_0^{\alpha\beta}({\bf q}) = 
-\frac1N \sum_{\bf k} \sum_{\eta\nu}
\frac{f_{\eta}({\bf k}) - f_{\nu}({\bf k}+{\bf q}) }
{\epsilon_{\eta}({\bf k}) - \epsilon_{\nu}({\bf k}+{\bf q}) }\times
\nonumber \\
{\cal S}^*_{\alpha\eta}({\bf k})
{\cal S}_{\alpha\nu}({\bf k}+{\bf q})
{\cal S}_{\beta\eta}({\bf k})
{\cal S}^*_{\beta\nu}({\bf k}+{\bf q})
\label{eq:rpa}
\end{eqnarray}
where ${\cal S}_{\eta\nu}({\bf k})$ are the similarity
transformations which diagonalize the $4\times4$ matrix defining the band
structure for each momentum ${\bf k}$.  Since the full susceptibility, within
the RPA, is given by $\chi = \chi_0 / ( 1 - U \chi_0)$, a magnetic phase
transition to a state with ordering vector ${\bf q}$ occurs when $U \chi_0$
reaches unity (Stoner criteron) at $U=U_c$. The RPA is exact in the limit 
$U\to0$, and is expected to be reasonably accurate at weak coupling. We carry 
out the calculation of $\chi_0$ with lattice sizes up to $40,000^2$ and at 
various low temperatures to be able to obtain results in the thermodyanamic 
limit at $T=0$. 

For every parameter set, the dominant magnetic order is the one with the 
smallest $U_c$. There are other, subleading, 
ordering modes with larger values of $U_c$. These modes do not, however,
replace the leading order when their respective Stoner criteria are
met, i.e., the simplest version of the RPA is no longer valid after the system 
enters a broken symmetry state. 

Unfortunately, away from half-filling, DQMC cannot reach low temperatures
due to the sign problem,
so, we cannot compare the RPA calculations with this exact approach.
Different vertex corrections can, however, be evaluated diagrammatically~\cite{vertex}.
They are not expected to change the dominant ordering wave vector in an
unfrustrated system with a few discrete ordering tendencies, like the one
studied here, although they change $U_c$. Hence, we expect the
vertex corrections to modify our RPA magnetic phase boundaries away from half
filling only by shifting the quantitate value of $U_c(\rho)$, but not the
symmetry of the ordered phase that appears for $U>U_c$. The effect of
various types of vertex corrections on these boundaries, and the role they
may play in studying pairing and superconductivity due to spin fluctuations
by diagrammatic methods on this geometry would be interesting problems for future research.

\end{document}